# Meta-morphism: Exotic Polymorphism of Metamaterial Self-assembled by pyrene derivative

**Kyoung Hwan Choi[1], Da Young Hwang[1], and Dong Hack Suh[1†]**

*1 Advanced Materials & Chemical Engineering Building 311, 222 Wangsimni-ro, Seongdong-Gu, Seoul, Korea, E-mail: dhsuh@hanyang.ac.kr]=*

**Abstract**

Polymorphism, which describes the occurrence of different lattice structures in a crystalline material, is a critical phenomenon in material science and condensed matter physics. It has emerged as a major focus for industry and regulatory agencies respectively. Thermomicroscopy, infrared spectroscopy and thermal analysis, especially differential scanning calorimetry (DSC) is used to characterize polymorphism to provide a powerful to isolate and identify of crystalline modification. Enantiotropic and monotropic with reversible endothermic and irreversible exothermic phase transition is representative classifications of polymorphism. Recently, Dirac metamaterial based on pyrene derivatives is attracting great attention. It succeeded in forming a periodic and regular structure using the unique π-π interaction of the pyrene derivative, namely HYLION-12. The phase transition between modifications is not classified into the existing polymorphism system. Here, we propose a new kind of polymorphism by identifying and analyzing thermodynamic functions such as heat capacity, enthalpy, entropy and, Gibbs free energy between modifications from DSC. This not only allows us to better understand the formation of Dirac materials at the molecular level, but also to think about the condition for new types of polymorphism.

**Introduction**

Metamaterials, artificial composite structures with exotic material properties, have emerged as a new frontier of science involving physics, material science, engineering and chemistry.[1,2] Metamaterials consist of periodically or randomly distributed structured elements, whose size and spacing are much smaller than the wavelength of electromagnetic wave.[3] The metamaterials paradigm has mostly been considered as a means of engineering the electromagnetic response of passive structured materials by engaging resonance excitations such as localized plasmonic modes.[4] Under the basic tenet of metamaterials that the physics is underpinned to the architecture, a unique type of organic crystal of HYLION-12 has been proposed with its novel periodicity and regularity compared to the existing molecular crystals. Apart from the exotic characteristics, HYLION-12 is yet hampered by the densest state of crystal structure which can be observed in conventional molecular crystals.

Depending on the formation of various crystal structures from the same molecule, polymorphism has expanded theoretical approaches since understanding the structural factor of properties is one crucial aspect in material design.[5] Polymorphism is also centered on the identification of structure-property relationships,[6] and its influence was thereby developed in the relevant fields once the property is varied by its crystal structures; i.e., pharmaceutical[7,8], dye[9] and explosives[10].

The thermodynamic relationships between modifications that make up polymorphism have been demonstrated and divided into two categories, monotropic and enantiotropic.[11-13] Monotropic involves the irreversible phase transition with exothermic heat exchange between the modifications. Contrastively, the enantiotropic implies the reversible phase transition with endothermic heat exchange.[14]

Here, we demonstrate the unique polymorphic characteristics of the metamaterial through polymorphism of HYLION-12. The phase transition between the orthorhombic phase ($\Phi_o$) with the metamaterial property and the monoclinic phase($\Phi_m$) with general organic crystals are not categorized by neither enantiotropic nor monotropic, given that it has irreversible endothermic phase transition from $\Phi_o$ to $\Phi_m$. Heat capacity, enthalpy, entropy and Gibbs free energy for each modification was explored using a thermodynamic approach with a difference scanning calorimeter (DSC) to introduce a novel kind of polymorphism. Further, different states of Gibbs free energy between modifications were calculated using solubility measurements for an advanced representation. The new class of approach here will provide a higher level of abstraction of the physical nature of polymorphism.

**Results and discussion**

HYLION-12 is the basic ingredient to form a metamaterial. Herein, crystals exhibit different characteristics apart from those of organic molecules, because of their unusual periodic and regular arrangement originated from a dimer of HYLION-12. On the basis, it is sufficient to form octahedral 6 coordination (i.e., transition metals) without being organic molecules.

In the case of HYLION-12, $\Phi_o$ is different from $\Phi_m$. For instance, $\Phi_o$ has a plenty of internal space to form the periodic structure, thus the state is not highly stable in the thermodynamic sense, while $\Phi_m$ reveal the most stable characteristics. There are numerous reports identifying their individual crystal structures. Thus far, however, no comprehensive research has clearly identified the polymorphism between $\Phi_o$ and $\Phi_m$.

In accordance with existing research, DSC was carried out to confirm the relationship between modifications.[15] $\Phi_o$ is used as the starting material for the DSC experiment. Two endothermic peaks were observed during the first heating cycle. The first peak at 340.62 K corresponds to the phase transition from $\Phi_o$ to $\Phi_m$ with an associated energy of 35.04 J/g. The second peak at 350.00 K corresponds to the melting point of HYLION-12 with an energy of 111.8 J/g (Fig. 1a). In the first cooling cycle, HYLION-12 crystallized at 335.36 K with an energy of 116.4 J/g (Fig. 1b). Only one endothermic peak was observed during the second heating cycle at

349.96 K with an energy of 108.3 J/g. This peak corresponded to the second peak in first heating cycle and indicates the phase transition from $\Phi_o$ to $\Phi_m$ is irreversible. (Fig. 1c). This is a peculiar from a normal soild-to-soild phase transition since the phase transition from an unstable phase to a stable phase is usually exothermic.

In terms of thermodynamics, polymorphic behaviors are classified as monotropic or enantiomeric.[11] Four criteria distinguish enantiotropic and monotropic systems: the heat of transition rule, the heat of fusion rule, the infrared rule and density rule. According to these rules, a monotropic has the irreversible exothermic phase transition, while an enantiotropic has a reversible endothermic phase transition.[11,16-18] However, $\Phi_o$ has an irreversible endothermic phase transition to $\Phi_m$. It is a type of the phase transition that is not categorized to either. Functionalizing the thermal behavior of each modification may bring more insight into the understanding of the exotic polymorphism.

**Heat capacity and enthalpy**

Among the four criteria of polymorphism, the density rule is easy to be decided. The density of each modification can be calculated from the volume of unit cell and total weight of atoms in unit cell. The resultant densities of $\Phi_o$ and $\Phi_m$ are 0.260g/cm³ and 1.502g/cm³, respectively. As expected, $\Phi_m$ is more stable at 0 kelvin.

To confirm the suitability of the other rules, the existing method was employed for the quantitative study of the heat capacity.[19] The heat capacity function of each modification in temperature ranges from 273K to 325K can be expressed by empirical equations (1) and (2):

$$\boldsymbol{C_{p,ortho}(T) = -194.048 + (0.570)T + 4.256 \times 10^6 \times T^{-2}} - (1)$$

$$\boldsymbol{C_{p,mono}(T) = -221.533 + (0.634)T + 5.014 \times 10^6 \times T^{-2}} - (2)$$

Results clearly confirm that the heat capacity of each modification by temperature has a non-linear relationship rather than a simple linear relationship. (Fig. 2a and b) In classical physics, heat capacity was considered constant according to the Dulong-Petit law, regardless of the type of materials.[20] However, Einstein and Debye showed that heat capacity is a function of frequency and temperature.[21] The energy-temperature diagram of polymorphism followed the Einstein and Debye model; thus far, however, the temperature has been solely considered as a variant of the heat capacity function because the frequency of molecular motion is not so different for various crystal structures. Even if there is a difference in frequency, the less dense modification is assumed to have the higher frequency.[22] According to these considerations, the less dense modification with a higher heat capacity than the denser modification has a relatively lower heat capacity.

Interestingly, $\Phi_o$ has a lower increment of heat capacity and vice versa in $\Phi_m$ as to the restriction of HYLION-12 toward the molecular motion to create regular and periodic structures in $\Phi_o$. The π-π interactions lock up the vibration of perpendicularly stacked pyrene along the z-axis and the alkyl chain overlaps restrict free motion of it.

The enthalpy curve would be obtained from integration of the heat capacity function.[23] The integral constant can be inferred from the endothermic heat exchange during the phase transition at 345.15K. According to the enthalpy differences, the enthalpy equations of both modifications are represented in following equations:

$$\boldsymbol{H_{ortho}(T) = -194.0481T + (0.28521)T^2 - 4.256 \times 10^6 \times T^{-1}} - (3)$$

$$\boldsymbol{H_{mono}(T) = -221.53315T + (0.31712)T^2 - 5.014 \times 10^6 \times T^{-1}} - (4)$$

The enthalpy curve of the modifications intersects at 207.8K. (Figure 2c) This well represents difference of polymorphism of the metamaterial from usual phenomena. In research of polymorphism; the enthalpy curves of the less dense phase and the denser phase increased with temperature can be viewed as almost parallel.[24] the reason is that the difference of frequency between modifications was not large enough. Therefore, only temperature is a dominant factor influencing enthalpy change. In $\Phi_o$, HYLION-12 is heavily constrained by its movement at the molecular level in order to maintain the structure. In addition, entwined HYLION-12 required high frequency to vibrate. The crossing of the enthalpy curve also represents the endothermic phase transition between modifications of HYLION-12.

**Reversibility**

On the other hand, reversibility between modifications is also a very important factor in research of polymorphism. As mentioned earlier, not only the relationship between heat exchanges, but also the reversibility is considered important to classify a category of polymorphism.[11,16-18] Irreversible phase transitions commonly occur when Gibbs free energy curves of each modification do not directly cross. On the other hands, reversible phase transitions occurred when the Gibbs free energy curves intersected.[25] In order to verify the reversibility between modifications, the entropy and Gibbs free energy functions are calculated from heat capacity functions with reference temperature as 1K.[26] As the results, entropy functions for each modification can be obtained as equations (5) and (6).

$$S_{ortho}(T) = (-194.058 + (0.570)T + 4.256 \times 10^6 \times T^{-2}) \ln T - (5)$$

$$S_{mono}(T) = (-221.533 + (0.634)T + 5.014 \times 10^6 \times T^{-2}) \ln T - (6)$$

The entropy of each modification sharply increases at low temperature and gradually approaches to zero with the positive values. (Figure 2d) The difference of entropy between modifications provides the significant clue of irreversibility. The entropy curves are nearly in contact, and the difference between modifications is only 0.2077 kJmol$^{-1}$ K$^{-1}$ at 287K.

According to the concept of the order-disorder transition, the less dense phase usually has the higher entropy.[11] Because the interactions between them would be weaker, they lead to a smaller binding energy. However, the entropy difference resulting from disorder is roughly

R ln (N) per mole, where N is the number of the equivalent position or orientations occupied at random by the molecules.[27]

The enthalpy and entropy functions can be used to determine the Gibbs free energy for each modification, and they are represented by equations (7) and (8):

$$\boldsymbol{G_{ortho}(T) = \left(-194.048T + (0.285)T^2 - \left(4.256 \times 10^6\right)T^{-1}\right)}$$
$$\boldsymbol{-T((-194.048 + (0.570)T + 4.256 \times 10^6 \times T^{-2})\ln T) - (7)}$$

$$\boldsymbol{G_{mono}(T) = \left(-221.533T + (0.317)T^2 - \left(5.014 \times 10^6\right)T^{-1}\right)}$$
$$\boldsymbol{-T((-221.533 + (0.634)T + 5.014 \times 10^6 \times T^{-2})\ln T) - (8)}$$

The Gibbs free energy for each modification demonstrate rational interpretation for the reason of exotic polymorphism from the metamaterial. (Figure. 2f) The Gibbs free energy of each modification is nearly in contact at 278K and the difference of Gibbs free energy is only 0.267. A novel type of phase transition that has never been confirmed before is observed in modifications of the metamaterial. (Figure. 2f inset) Crossed enthalpy and contacted Gibbs free energy curves provide with the only way to explain the contradictory phase transition of the metamaterial accompanied with irreversibility and endotherm. The Gibbs free energy curves of traditional polymorphism have been crossed or paralleled so far. It confirms why the phase transition of modifications of HYLION-12 cannot be categorized with a conventional polymorphism.

A more direct demonstration of the Gibbs free energy relation between modifications was made through the solubility of two modifications. The solubility of each modification was investigated for a more precise and detailed analysis of this phenomenon because the differences in solubility were closely related to the change in Gibbs free energy. The difference in Gibbs free energy based on the solubility was used to solve the problems that can be addressed from this assumption. The differences in solubility between the modifications of

HYLION-12 are useful for confirming the curve contact of Gibbs free energy.[28,29] The change in Gibbs free energy with respect to solubility can be represented by the equation (9).

$$\Delta G = RT \ln \frac{L_{ortho}}{L_{mono}} \quad (9)$$

In the contact model of Gibbs free energy, Gibbs free energy of $\Phi_o$ is always larger than that of $\Phi_m$. Therefore, before contact temperature, the differences in Gibbs free energy between the two modifications decrease as temperature increases. After the contact of Gibbs free energy, the differences increase with increasing temperature.

The experiment was based on the Beer-Lambert law.[30] Transmittance for various concentrations of HYLION-12 was measured and calculated in UV in order to determine the absorption coefficient at 348nm. (Figure. 3a and 3b) To confirm the solubility of each modification, the same weight of each modification was dissolved to the 5ml $CHCl_3$ at various temperature and absorption was measured (Figures 3c and 3d). The solubility of each modification represents a good linearity when expressed as a Van't hoff plot and solubility of each modification is expressed as following equations (10) and (11):

$$L_{mono}(T) = -12771.27 \frac{1}{T} + 37.16054 \quad (10)$$

$$L_{ortho}(T) = -4712.93 \frac{1}{T} + 10.79747 \quad (11)$$

The results show that solubility of $\Phi_m$ has a lower intercept than that of $\Phi_o$ and increases rapidly. **(Figure. 3e)** Solubility experiments can indirectly confirm the contact observations for the Gibbs free energy curves. By assigning $L_{ortho}$ and $L_{mono}$ to equation (9). It distinctly shows the expected pattern with a minimum value at 284K. **(Figure 3f)** It is obtained as a very similar result from the thermodynamic calculation.

**Conclusion**

We have discovered a novel type of polymorphism with an irreversible endothermic phase transition, called meta-morphism. Major characteristics were observed in Gibbs free energy-temperature diagrams with the enthalpy curve crossing and the contact of Gibbs free energy between modifications. The phenomenon can be explained by the heat capacity differences of modifications, which are caused by the restriction of molecular motion of the less dense phase. The key reason to this exotic polymorphism is the restriction of frequency among molecules. Evidently, meta-morphism is a new interpretation and discovery for thermodynamics of the molecular metamaterial, which is useful in not only understanding polymorphisms and phase transitions, but also crystal engineering for the structural design of the dynamic phase transition.

Figures

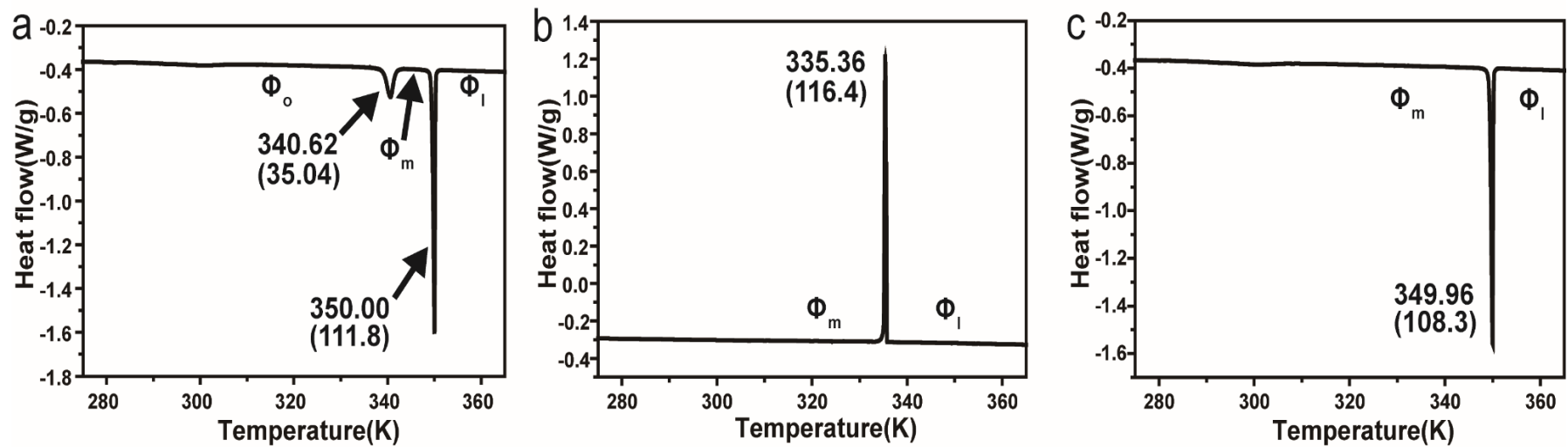

**Figure 1. Differential scanning calorimetry of HYLION crystals. a.** The first heating cycle. The first phase transition shows the phase transition from orthorhombic to monoclinic and the second phase transition from monoclinic to liquid; **b.** The first cooling cycle: the crystallization of the monoclinic phase **c**. The second heating cycle: disappearance of the unstable orthorhombic phase.

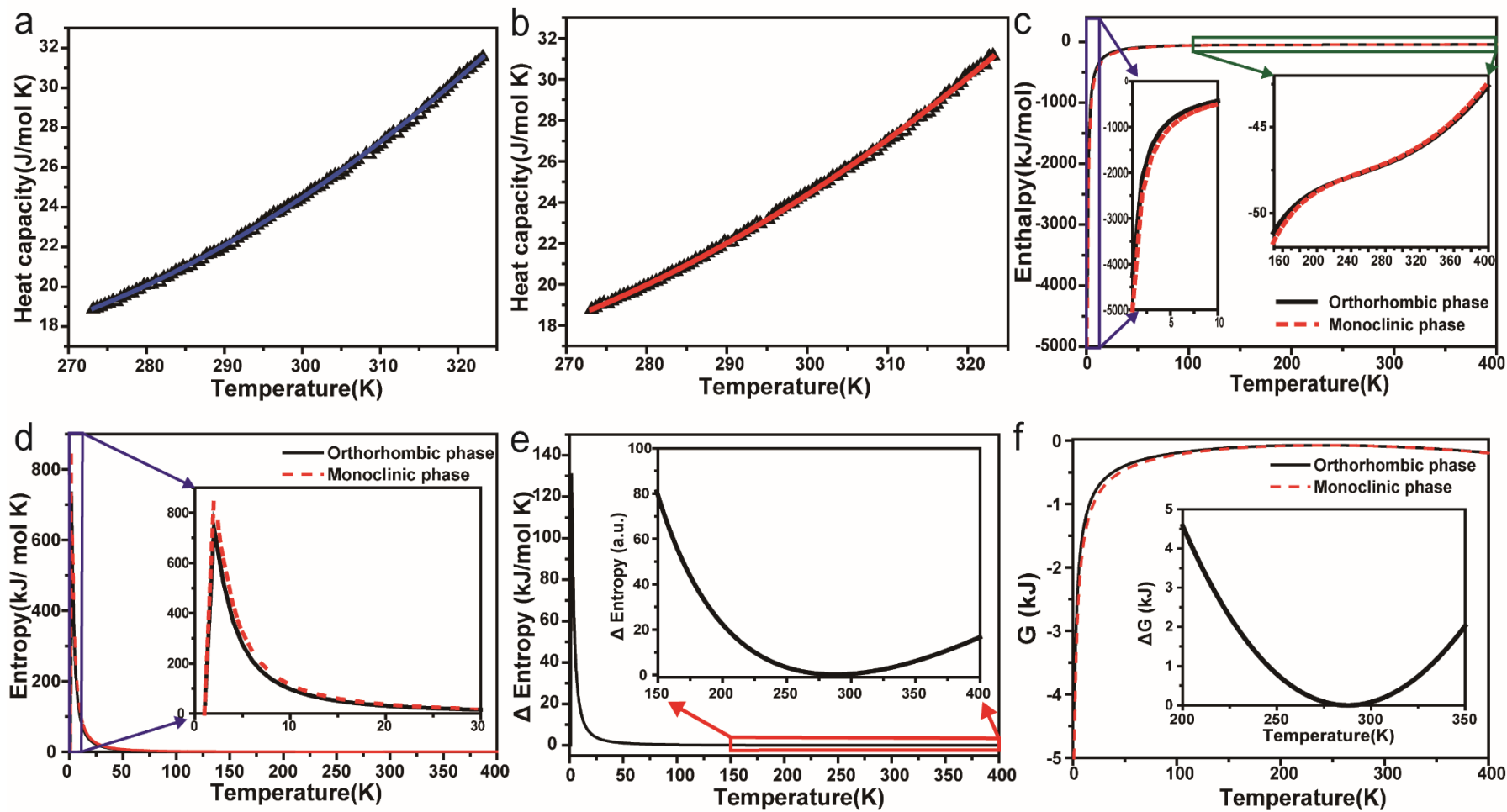

**Figure 2**. **Thermodynamic functions of Dirac metamaterial of HYLION-12. a and b.** Heat capacity versus temperature diagrams for each modification of HYLION-12. **c and d.** Enthalpy vs. temperature, and entropy vs. temperature diagrams for each modification, orthorhombic (black) and monoclinic (red dash). **e.** Difference of entropy between modifications. Inset clearly shows the difference of entropy with nearly zero at 287K **f.** Gibbs free energy of each modification. Inset is difference of Gibbs free energy between modifications and it is about 0.267 at 287K.

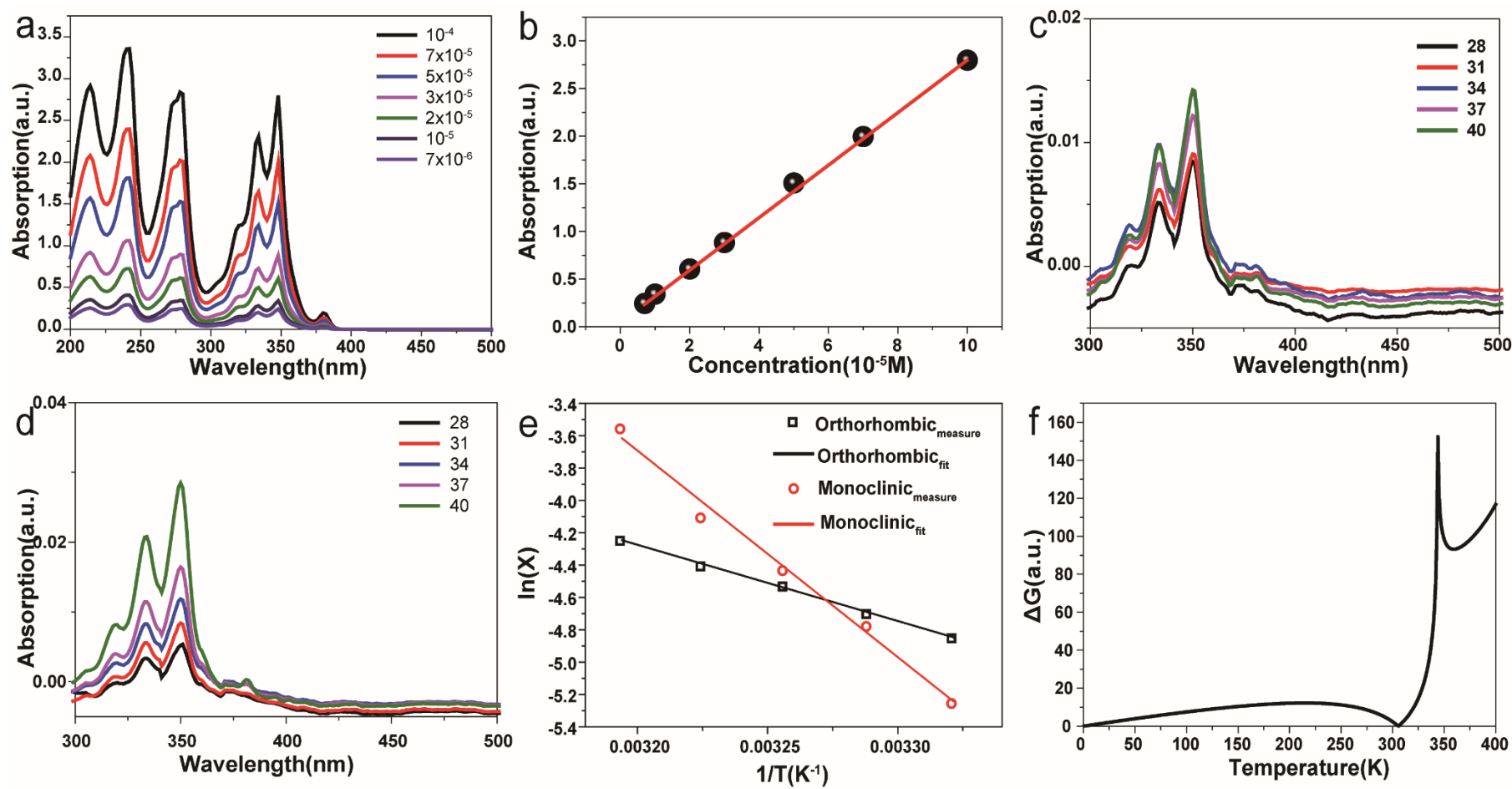

**Figure 3**. **UV/vis spectroscopy to verify the solubility of modifications. a.** The UV/vis absorption spectra of diluted solution of HYLION-12 in CHCl$_3$. **b.** Concentration versus absorption. Linear relationship between concentration and absorption can determine the molecular extinction coefficient of HYLION-12. **c and d.** The UV/vis absorption spectra of saturated solution represented by orthorhombic and monoclinic phases in various temperatures. **e.** Van't Hoff plots for each modification. **f.** Differences of Gibbs energy between modifications calculated from solubility.

**Methods**

**General procedures** Pyrene, $NaIO_4$, $Bu_4NBr$, $Na_2S_2O_4$, and dodecyl bromide were purchased from Sigma Aldrich and used as received. $CDCl_3$ was purchased from Cambridge Isotope Laboratories and used for the $^1H$ NMR spectroscopic studies.

**Synthesis and crystallization** A mixture of pyrene-4,5,9,10-tetrone (10 mmol), $Bu_4NBr$ (13 mmol), and $Na_2S_2O_4$ (115 mmol) in $H_2O$ (50 ml) and THF (50 ml) was shaken for 5 min. The color of the mixture changed from dark brown to pale yellow. Dodecyl bromide (60 mmol) was added, followed by aqueous KOH (306 mmol, in 50 ml $H_2O$). The mixture was stirred overnight and poured into $H_2O$ (50 ml) and EA (30 ml). The yellow solid was filtered and washed with EtOH. After drying in a vacuum, the solid was recrystallized from EA, resulting in a white solid of yield 82% to 85%. To prepare the sample for powder X-ray diffraction (PXRD), 4,5,9,10-tetrakis(dodecyloxy)-pyrene was crystallized through a composition gradient method from n-hexane of concentrations less than $10^{-3}$ M.

**Characterization**

$^1H$ NMR (600 MHz) spectra were recorded on a Bruker DRX VNMRS 600 instrument. The purity of the products was determined by a combination of thin-layer chromatography (TLC) on silica gel coated aluminum plates (with an F254 indicator; layer thickness 200 μm; particle size 2-25 μm; pore size 60Å, SIGMA-Aldrich). Infrared spectra were taken in a KBr disc using a Jasco FT-IR 460 plus spectrometer. Powder X-ray diffraction patterns of the samples were measured on a Brucker D8 advance and Rigaku D/MAX RINT 2000 diffractometer using monochromatized Cu-Kα (λ = 0.15418 nm) radiation at 40 kV and 100 mA. The thermal transition was measured on a TA Instrument SDT Q 600 ver. 20.9 Build 20 system with differential scanning calorimetry (DSC) integrated with a refrigerated cooling system (RCS). In all of the cases, the heating and cooling rates were 1 K min$^{-1}$. The transition temperatures

were measured as the maxima and minima of their endothermic and exothermic peaks. MATLAB R2012b (8.0.0.783) was used for plotting the graph.

**Supporting information**

**Meta-morphism: Exotic Polymorphism of Metamaterial self-assembled by pyrene derivative**

| Equation | $y = A + B*T + C*T^{-2}$ | | |
|---|---|---|---|
| Reduced Chi-Sqr | 0.00801 | | |
| Adj. R-Square | 0.9994 | | |
| | | Value | Standard Error |
| A | | -194.048 | 4.17839 |
| B | | 0.57042 | 0.00936 |
| C | | 4.26E+06 | 122693 |

**Table S1.** Fitting results of heat capacity for orthorhombic phase of HYLION-12

| Equation | $y = A + B*T + C*T^{-2}$ | | |
|---|---|---|---|
| Reduced Chi-Sqr | 0.00652 | | |
| Adj. R-Square | 0.99953 | | |
| | | Value | Standard Error |
| A | | -221.53315 | 3.75009 |
| B | | 0.63425 | 0.0084 |
| C | | 5.014E+06 | 110147 |

**Table S2. Fitting results of heat capacity for monoclinic phase of HYLION-12**

| Equation | y = a + b*x | Pearson's r | 0.99952 |
|---|---|---|---|
| Residual Sum of Squares | 0.00518 | Adj. R-Square | 0.99884 |
| | | Value | Standard Error |
| | Intercept | 0.06659 | 0.01985 |
| | Slope | 27548.78848 | 382.49065 |

**Table S3.** Fitting results of absorption coefficient of HYLION-12

| Equation | y = A*T$^{-1}$ + B | | | | |
|---|---|---|---|---|---|
| | Reduced Chi-Sqr | Adj. R-Square | Coefficient | Value | Standard Error |
| Orthorhombic | 1.10148x10$^{-4}$ | 0.99804 | A | -4712.92506 | 104.32714 |
| | | | B | 10.79747 | 0.33976 |
| Monoclinic | 0.00509 | 0.98778 | A | -12771.27102 | 709.07665 |
| | | | B | 37.16054 | 2.30923 |

**Table S4.** Fitting results of Van't Hoff plots for each modification